\documentclass[aps,prd,reprint,superscriptaddress,nofootinbib]{revtex4-1}
\usepackage{CJK}

\usepackage{amsmath,bm,bbold}
\usepackage{graphicx}
\usepackage[braket,qm]{qcircuit}
\usepackage{multirow}
\usepackage[colorlinks,citecolor=blue]{hyperref}

\usepackage[dvipsnames,usenames]{xcolor}

\begin{document}
\begin{CJK*}{UTF8}{gbsn}
\title{Many-body effects of collective neutrino oscillations}
\author{Zewei Xiong (熊泽玮)}
\affiliation{GSI Helmholtzzentrum f{\"u}r Schwerioneneforschung, 64291 Darmstadt, Germany}
\date{\today}

\begin{abstract}
Collective neutrino oscillations are critical to determine the neutrino flavor content, which has striking impacts on core-collapse supernovae or compact binary merger remnants.
It is a challenging many-body problem that so far has been mainly studied at the mean-field approximation.
We use a setup that captures the relevant physics and allows exact solution for a large number of neutrinos.
We find that quantitative deviation from the mean-field evolution can exist even for a large system.
The underlying mechanism due to many-body decoherence in flavor space is analyzed, and similar features have been observed in a spin-1 Bose-Einstein condensate.
Our results call for more careful examinations on the possible many-body corrections to collective neutrino oscillations in astrophysical environments.
\end{abstract}

%\keywords{Suggested keywords} % Use show keys class option if keyword
\maketitle
\end{CJK*}

\section{Introduction}
Dense astrophysical environments such as core-collapse supernovae and neutron-star mergers provide unique laboratories to probe rich phenomena of neutrino oscillations and its flavor evolution.
The neutrino flavor content determines the impact of charged-current reactions as they mainly involve $\nu_e$ and $\bar\nu_e$ in the astrophysical environment.
Large neutrino fluxes emitted from the protoneutron star, in the case of supernovae, or from the hypermassive neutron star remnant, in the case of mergers, not only provide heating and cooling mechanisms on the ejecta, but also alter the nucleosynthesis by determining the neutron-to-proton ratio~\cite{frohlich2006neutrino,duan2011influence,wanajo2013r,wu2015effects,sasaki2017possible,wu2017imprints,stapleford2020coupling,xiong2020potential,george2020fast,li2021neutrino}.
Additionally, the flavor ratio is a key observable for the next Galactic supernova event~\cite{mirizzi2016supernova,guo2019presupernova,hansen2020timing,li2021exciting}.

Neutrino oscillations in vacuum and ordinary matter are well established by experiments with solar, atmospheric, reactor, and accelerator~\cite{olive2014review}.
In addition, neutrinos propagating in a dense neutrino gas undergo coherent forward scattering among each other. 
This refractive effect leads to various behaviors of collective flavor oscillations including slow flavor instability~\cite{pastor2002physics,duan2007picture,raffelt2007self,duan2010collective,chakraborty2016collective}, fast flavor instability~\cite{sawyer2005speed,izaguirre2017fast,yi2019dispersion,johns2020neutrino,tamborra2020new,xiong2021stationary,padilla2021neutrino,dasgupta2017fast,abbar2019occurrence,azari2020fast,morinaga2020fast}, and matter-neutrino resonance~\cite{malkus2012neutrino,wu2016physics,zhu2016matter}.

While most of the above manifestations have been studied at the mean-field approximation, we know that neutrinos constitute a many-body quantum gas affected by many-body entanglement.
Exact solutions involving few neutrinos have become available in the past decade~\cite{bell2003speed,friedland2003neutrino,sawyer2004classical,cervia2019entanglement,rrapaj2020exact}.
The main obstacle placed on those studies is, however, the drastically increasing many-body Hilbert state space for a multi-spin-like system with total particle number $N$, which goes as $\sim \mathcal O(e^N)$, considering that we are more interested in the time evolution of the whole system rather than the behaviors of the ground state and few excited states.
A variety of efforts have been devoted to surmount this exponential barrier~\cite{pehlivan2011invariants,birol2018neutrino,patwardhan2019eigenvalues,patwardhan2021spectral}.
Those approaches provide encouraging progress, but none of them at the moment shows a convergence to the properties of the infinite system.
This drawback could be due to the limited number of neutrinos considered: $N\lesssim 20$.
Tensor network methods~\cite{roggero2021entanglement,roggero2021dynamical} can provide an alternative to study larger-scale systems and understand the thermodynamic limit; however, at the moment they are still limited to $\sim \mathcal O(100)$ neutrinos.
The analytical scheme of simplified oscillation model based on the analogous of angular momentum representation in Refs.~\cite{friedland2003many,friedland2006construction} is able to handle a large system, but it is restricted to a two-beam setup, i.e. neutrinos moving in two different directions, without the one-body vacuum term.
In this setup they can take advantage of the fact that the total projected flavor isospin commutates with the Hamiltonian and solve the evolution equation analytically, but it does not allow any mean-field flavor instability with exponentially growing modes so only flavor conversion occurring in a time scale of $\sim \mathcal O(\sqrt{N})$ was discovered.

Following those pioneering works, we propose a new method that generalizes the angular momentum scheme in a simple and efficient numerical way for setups that allow exponentially growing flavor instabilities in the mean-field level.
We mainly focus on two-beam slow flavor instability with zero vacuum mixing angle and show that the flavor evolution qualitatively converges to a bipolar motion.
This two-beam setup is the minimal nontrivial configuration that captures the property of bipolar motion, which is the crux of collective neutrino flavor instabilities.

We reveal a many-body effect leading to decoherence in the flavor evolution that quantitatively deviates from mean-field predictions even for a large system with the number of neutrinos $\sim \mathcal O(10^6)$.
We discuss an extension to more beams and nonzero vacuum mixing angle for other flavor instabilities.
We also point out the similar features shared with other many-body systems and discuss the potential experimental explorations.

\section{Method}
Although collective neutrino oscillations are three-flavor phenomena~\cite{duan2008stepwise,dasgupta2008spectral,capozzi2020mu,shalgar2021three}, we consider, for simplicity, two neutrino flavors: electron neutrinos $\nu_e$ and heavy-lepton neutrinos $\nu_x$.
The Hamiltonian describing oscillations is 
\begin{equation}
    H 
    = \sum_{k=1}^N \frac{\omega_k \bm{\mathcal{B}} \cdot \bm{\tau}_k + \lambda \tau_k^z }{2} +
    \sum_{i<j}^N \frac{\mu J_{ij}}{2N} (\bm{\tau}_i \cdot \bm{\tau}_j + \mathbb 1_i \mathbb 1_j),
    \label{eq:EOM_manybody}
\end{equation}
where $\omega_k=\Delta m^2/2E_k$ is the vacuum oscillation frequency in terms of mass-square difference $\Delta m^2$ and energy $E_k$ for the $k$th neutrino,
$\bm{\tau}$ is the Pauli matrix operating in flavor space, $\mathbb 1$ is an identity matrix,
$\lambda = \sqrt{2} G_F n_e$ is the effective potential with matter, $G_F$ is the Fermi constant, $n_e$ is the net electron number density,
$\mu = \sqrt{2} G_F N/V$ represents the neutrino self-interaction strength, $N$ is the conserved neutrino number in a volume of $V$, and $J_{ij}$ is the two-body coupling coefficient between the $i$th and $j$th neutrinos.
The vacuum mixing vector $\bm{\mathcal{B}}$ is $(\sin 2\theta_V, 0, -\cos 2\theta_V)$ in normal mass hierarchy or $(-\sin 2\theta_V, 0, \cos 2\theta_V)$ in inverted mass hierarchy with the vacuum mixing angle $\theta_V$.
Despite the all-to-all interacting nature of the self-scattering term, this Hamiltonian is similar to a Heisenberg model.
The two-body operator in the self-scattering term can be rewritten as
\begin{equation}
    \bm{\tau}_i \cdot \bm{\tau}_j + \mathbb 1_i \mathbb 1_j =
    (\mathbb 1_i \mathbb 1_j + \tau^z_i \tau^z_j) + 2 \tau^+_i \tau^-_j + 2 \tau^-_i \tau^+_j,
    \label{eq:self_2body_operator}
\end{equation}
where $\tau^\pm=(\tau^x \pm i \tau^y)/2$ are the ladder operators.
The first two terms only contribute a phase, while the other two lead to the particle exchange between $\nu_e$ and $\nu_x$.

In our model, the neutrinos are divided into two beams with all $\nu_e$ in beam $A$ and $\nu_x$ in beam $B$ at the initial time.
Those neutrinos in the same beam move in almost the same velocity and share the same vacuum oscillation frequency $\omega_{I}$ where $I=A, B$.
The neutrino number of each beam $n_{I}$ is conserved and the initial total wave function is given as Slater determinant corresponding to the state with maximum flavor isospin.
As the system evolves, the flavor isospin for each beam is conserved and, hence, is convenient to work in a flavor isospin representation. 

We denote the total flavor isospin of both beams $T_A=n_A/2$ and $T_B=n_B/2$ as well as the projection $m_A=n_A/2$ and $m_B=-n_B/2$.
The total flavor isospin projection is defined as $M=m_A+m_B=(n_A-n_B)/2$.
At large matter densities, the effective mixing angle is highly suppressed, so the first component of $\bm{\mathcal{B}}$ is negligible.
Assuming inverse mass hierarchy, the Hamiltonian becomes
\begin{equation}
    H = \frac{\omega_A' \tau^z_A + \omega_B' \tau^z_B}{2} +
    \frac{\mu J_{AB}}{2N} (\bm{\tau}_{A} \cdot \bm{\tau}_{B} + n_{A} n_{B}),
    \label{eq:EOM_manybody2_bybeam}
\end{equation}
where $\omega_{I}'=\omega_{I} + \lambda$ and $\bm{\tau}_{I} = \sum_{i\in I} \bm{\tau}_{i}$.
Given that the choice of beam velocities in two-beam model effectively changes only the self-interaction strength, we take $J_{AB}=1$ in the following discussion.

This Hamiltonian conserves the total isospin projection.
As a result, the evolving wave function is a linear combination of many-body states with all possible $\nu_e$-$\nu_x$ pair-exchange numbers $p$, such as
$m_A=n_A/2-p$ and $m_B=p-n_B/2$.
Hence we can characterize the states by the number $p$: $\ket{p}\equiv \ket{T_A, m_A} \otimes \ket{T_B, m_B}$; and the wave function is written as $ \ket{\Psi(t)} = \sum_{p=0}^{p_{\mathrm{max}}} a_p(t) \ket{p}$, with the maximal pair-exchange number $p_{\text{max}} = \mathrm{min}(n_A, n_B)$ and the time-dependent amplitude $a_p(t)$.

% \paragraph{Equations of motion.}
Since the Hamiltonian of Eq.~(\ref{eq:EOM_manybody2_bybeam}) can connect only many-body states differing by zero or one pair-exchange number, the evolution is determined by solving the equations of motion (EOM): 
\begin{align}
    i \partial_t a_{p}
    & = i \partial_t \ip{p}{\Psi}
    = \sum_{p'=0}^{p_{\mathrm{max}}} \melem{p}{H}{p'}\ip{p'}{\Psi} \nonumber\\
    & = H_{p}^{p-1} a_{p-1} + (H_{p}^{(1)}+H_{p}^{p}) a_{p} + H_{p}^{p+1} a_{p+1},
    \label{eq:EOM_2beam}
\end{align}
where the contribution from the two-body operator is
\begin{align}
    H_{p}^{p-1}
    & = \frac{\mu}{N} \, p\sqrt{(n_A-p+1)(n_B-p+1)}, \nonumber\\
    H_{p}^{p} 
    & = \frac{\mu}{N} \, [p(n_A-p)+p(n_B-p)], \nonumber\\
    H_{p}^{p+1}
    & = \frac{\mu}{N} \, (p+1)\sqrt{(n_A-p)(n_B-p)},
    \label{eq:tranelem_2beam}
\end{align}
and the contribution from the one-body operator is
\begin{align}
    H_{p}^{(1)}
    & = - 2 \delta_\omega p +(n_A \omega_A'-n_B \omega_B')/2 ,
    \label{eq:tranelem_1body_2beam}
\end{align}
with $\delta_\omega=(\omega_A-\omega_B)/2$.
The second term in the one-body operator is a constant independent of $p$ that contributes only to the diagonal and hence does not affect the evolution.

Once the amplitudes for all many-body states are known, we can calculate the associated physical observables.
The averaged electron flavor fraction in each beam is determined as
\begin{align}
    \mathcal{P}_{A} & = \sum_{p=0}^{p_{\mathrm{max}}} \left(1-\frac{p}{n_A}\right) |a_{p}|^2, 
    \mathcal P_{B} = \sum_{p=0}^{p_{\mathrm{max}}} \frac{p}{n_B} |a_{p}|^2,
    \label{eq:P_2beam}
\end{align}
respectively.
To illustrate the difference between mean-field and the exact solution we use two different measures: the pair correlation and entanglement entropy.
We define the pair correlation for beam $A$ as
\begin{align}
    \mathcal C_{AA} & = 
    \melem{\Psi}{\left(\frac{\tau^z_A}{2 n_A}+\frac{1}{2}\right)^2}{\Psi} - \mathcal P_{A}^2 \nonumber\\
    & =
    - \mathcal P_{A}^2 + \sum_{p=0}^{p_{\mathrm{max}}} \left(1-\frac{p}{n_A}\right)^2 |a_{p}|^2.
    \label{eq:C_2beam}
\end{align}
and similar definitions for $\mathcal{C}_{BB}$ and $\mathcal{C}_{AB}$. 
The von Neumann entanglement entropy
\begin{equation}
    \mathcal S = - \sum_{p=0}^{p_{\mathrm{max}}} |a_p|^2 \log_2 |a_p|^2,
    \label{eq:von_neumann_entropy_gen}
\end{equation}
is independent of the beam. 

The mean-field approximation assumes negligible correlation between two neutrinos, i.e., $\langle \bm{\tau}_i \cdot \bm{\tau}_j \rangle = \langle \bm{\tau}_i \rangle \cdot \langle \bm{\tau}_j \rangle$.
Then, it is enough to evolve the single-particle density matrix for the $i$th neutrino, $\varrho^{(i)}$, using the mean-field EOM:
$i \partial_t \varrho^{(i)} = [H^{(i)}_{\mathrm{MF}}, \varrho^{(i)}]$,
where
$H^{(i)}_{\text{MF}} = (\omega_i+\lambda) \tau^z/2 + (\mu/N) \sum_{j\neq i}^N J_{ij} \varrho^{(j)} $.
For our two-beam setup, nonvanishing oscillations in the averaged flavor fraction, so-called flavor instabilities, appear whenever $\left( n_A/N - 1/2 \right)^2 + \left( \delta_\omega/\mu - 1/2 \right)^2 < 1/4$~\cite{duan2015flavor}.

\section{Many-body flavor evolution}
\begin{figure}[t]
	\centering
		\includegraphics[width=0.48\textwidth]{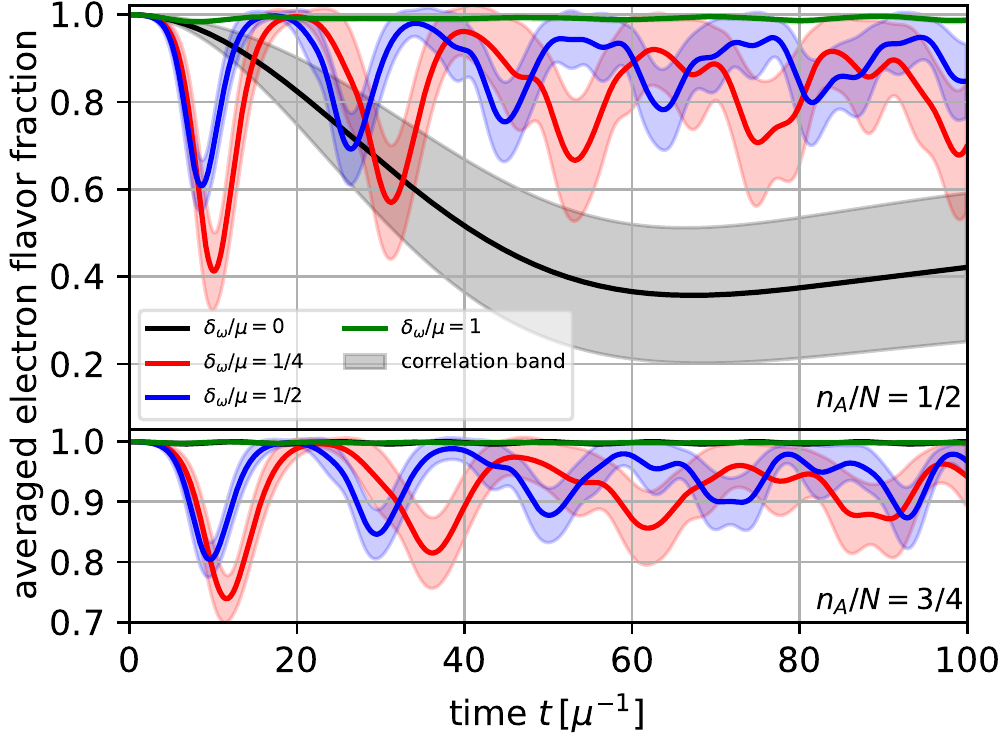}
	\caption{\label{fig:Pt_2beam} Evolution of averaged electron flavor fraction $\mathcal P_{A}$ (solid curves) for various $\delta_\omega/\mu$ and $n_A/N$ when $N=1000$. The light band of each color is plotted between $\mathcal P_A \pm \sqrt{\mathcal C_{AA}}/2$ to indicate the many-body uncertainty.}
\end{figure}

We solve the EOM in Eq.~(\ref{eq:EOM_2beam}) for four choices of $\delta_\omega/\mu=0$, $1/4$, $1/2$, and $1$ in equal-partition ($n_A=N/2$) and $\nu_e$-dominant ($n_A=3N/4$) cases respectively.
The total neutrino number is $N=1000$.
Figure~\ref{fig:Pt_2beam} shows the evolution of the averaged electron flavor fraction $\mathcal P_{A}$.
Consistent with the mean-field instability analysis, for both equal-partition and $\nu_e$-dominant cases, significant flavor oscillations occur for $\delta_\omega/\mu=1/4$ and $1/2$.
In the $\nu_e$-dominant case, the parameters $\delta_\omega/\mu=0$ and $1$ do not satisfy the unstable condition and there is almost no flavor transition.
The equal-partition case for $\delta_\omega/\mu=1$ is at the edge of instability region, so the flavor conversion is strongly suppressed and will further decrease with increasing system size.
An exception in equal-partition setup is that a large flavor transition is found when $\delta_\omega/\mu=0$.
This is the same instability reported in Ref.~\cite{friedland2003many} where the transition timescale was shown to scale with the number of neutrinos as $\sim \mathcal O(\sqrt{N})$.
Hence, it does not occur for a macroscopic neutrino gas.

The square root of pair correlation $\sqrt{\mathcal C_{AA}}$, i.e., the standard deviation of flavor fraction, is shown by the light bands around curves of $\mathcal P_A$ in Fig.~\ref{fig:Pt_2beam}.
The correlation remains zero when there is no flavor instability but can be gradually enhanced in unstable cases.

\begin{figure}[t]
	\centering
		\includegraphics[width=0.48\textwidth]{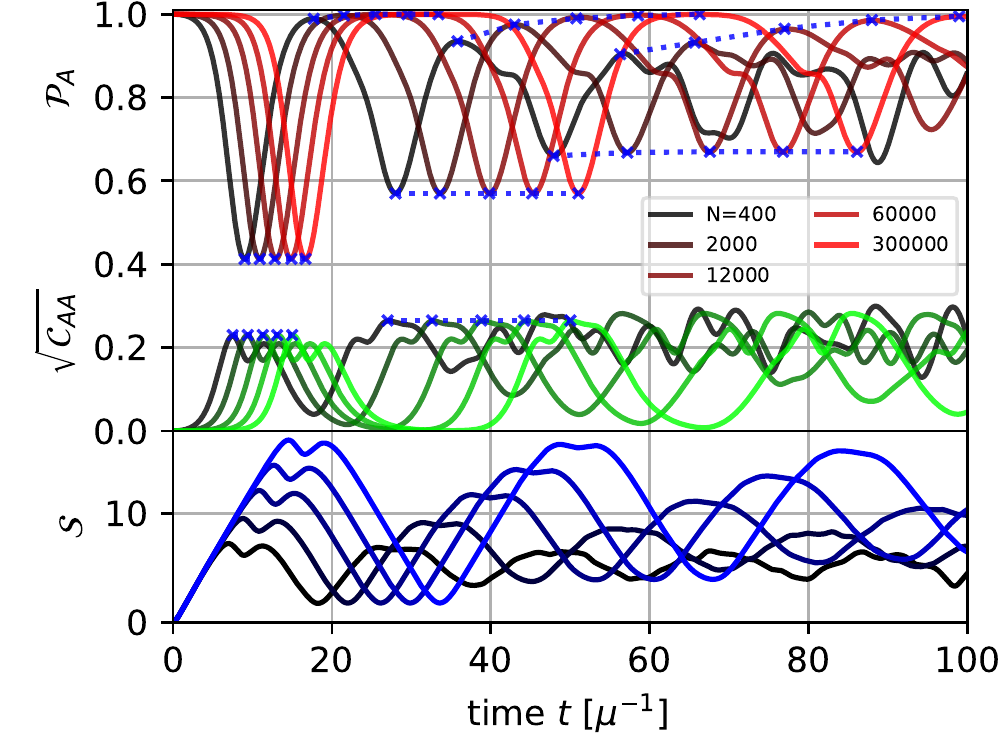}
	\caption{\label{fig:convergence_2beam} Evolution of $\mathcal{P}_{A}$ (red curves), $\sqrt{\mathcal{C}_{AA}}$ (green curves), and entanglement entropy $\mathcal S$ (blue curves) for different particle numbers $N=400-3\times 10^5$ (solid curves from dark to light). The minima and maxima of $\mathcal P_{A}$ in the first three cycles and maxima of $\sqrt{\mathcal{C}_{AA}}$ in the first two cycles for each case are labeled by blue cross markers and linked by blue dotted lines.
	}
\end{figure}

To explore the dependence with number of particles of the oscillation amplitude, its convergence and the behavior of the pair correlation, we examine cases with different numbers of neutrinos $N$ for $\delta_\omega/\mu=1/4$ and $n_A=N/2$.
The upper panel in Fig.~\ref{fig:convergence_2beam} shows $\mathcal{P}_{A}$ and $\sqrt{\mathcal{C}_{AA}}$ for $N=400$, $2000$, $1.2\times 10^4$, $6\times 10^4$, and $3\times 10^5$.
The flavor evolution follows a bipolarlike motion with a lower value for the averaged electron flavor fraction at each cycle that is independent of the number of particles considered and an upper value that approaches $\mathcal P_A= 1$ with increasing number of particles.
The maxima of the flavor transition probability are associated with maxima in the pair correlation and entanglement entropy.
The entanglement entropy reaches a maximal value slightly less than $\sim \log_2 N$, which is the maximal entropy with equal partition in $p_{\mathrm{max}}+1$ many-body states.
Although the flavor fraction is very close to one after each cycle, the entropy remains finite, and the system does not return back to its mean-field initial state.
The amount of disorder increases after each cycle.

\begin{figure}[t]
	\centering
		\includegraphics[width=0.45\textwidth]{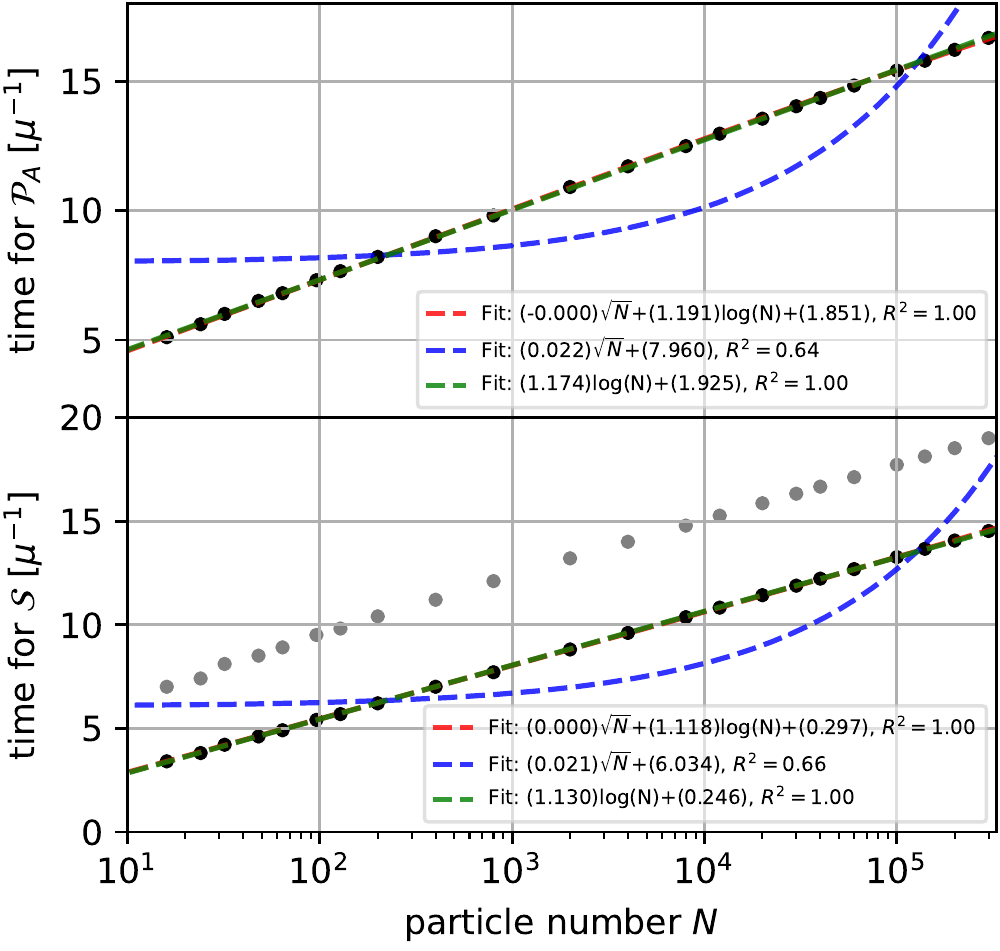}
	\caption{\label{fig:fit_2beam} The particle number dependence of the time to reach the first minimum of $\mathcal{P}_{A}$ (black dots in the upper panel), the first peak of $\mathcal{S}$ (black dots in the lower panel), and the second peak of $\mathcal{S}$ (gray dots in the lower panel). Black points correspond to $N=16$, 24, 32, 48, 64, 96, 128, 200, 400, 800, 2000, 4000, 8000, $1.2\times10^4$, $2\times10^4$, $3\times10^4$, $4\times10^4$, $6\times10^4$, $10^5$, $1.4\times10^5$, $2\times10^5$, and $3\times10^5$ respectively. Three functions of $\log N$ (green curves), $\sqrt{N}$ (blue curves), and their mixture (red curves) are used to fit the black dots in both panels. Best-fitted coefficients as well as the coefficient of determination $R^2$ are listed in the legend.}
\end{figure}

Another important characteristic is the timescale at which flavor instability develops.
We make function fitting on the timescales to reach first minimum of flavor fraction $\mathcal P_A$ and first maximum of entanglement entropy $\mathcal S$ with respect to the particle number in the system.
Figure~\ref{fig:fit_2beam} shows how the timescales grow as particle number $N$ increases as well as the fitting result.
Three functions are used: $f_1 = a\sqrt{N}+b\log N+c$, $f_2 = a\sqrt{N}+c$, and $f_3 = b\log N+c$.
The fitting result is strongly in favor of logarithmic dependence, consistent with the prediction from Ref.~\cite{roggero2021entanglement}.

\begin{figure}[t]
	\centering
		\includegraphics[width=0.48\textwidth]{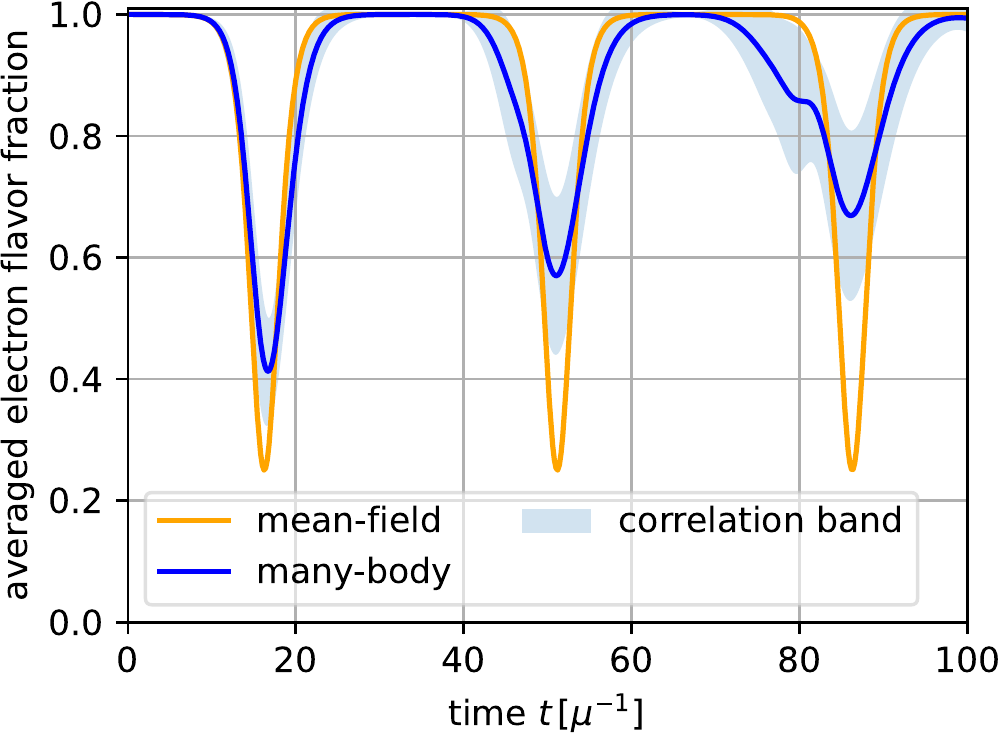}
	\caption{\label{fig:MF_MB_2beam} Comparison of $\mathcal{P}_{A}$ from mean-field evolution with the many-body result for $N=3\times 10^5$, $n_A=N/2$, and $\delta_\omega/\mu = 1/4$. Mean-field calculation is started with an initial perturbation of $\sim 10^{-3}$ on $\varrho_{ex}$. The light band is similar to that in Fig.~\ref{fig:Pt_2beam}.}
\end{figure}

This logarithmic dependence is associated with the quantum effect initiating the evolution in the absence of vacuum mixing in Refs.~\cite{sawyer2020quantum,sawyer2021neutrino}.
We compare in Fig.~\ref{fig:MF_MB_2beam} the evolution histories from both mean-field and many-body calculations for a large system of $N=3\times 10^5$.
A correlation band similar as in Fig.~\ref{fig:Pt_2beam} is plotted around the many-body result.
In mean-field calculation, we assign initial perturbations on the off-diagonal elements of density matrix $\varrho_{ex}=10^{-3}$ for neutrinos in beam $A$ and $\varrho_{ex}=\sqrt{3}\times 10^{-3} i$ in beam $B$ to trigger the flavor conversion.
The diagonal elements of the density matrix are adjusted accordingly for the normalization.
No artificial perturbation is assigned for the many-body case.
Both many-body and mean-field calculations show a similar exponential growth in the linear regime and a bipolar motion later, but the decoherence effects in the many-body calculation lead to a decrease in the amplitude and a growth in the entanglement.

The many-body effect to initiate the evolution can also be understood in terms of the quantum uncertainty of our initial flavor isospin state $\ket{T_A,m_A}$ for beam $A$.
This state has well-defined total flavor isospin $T^2 = n_A(n_A+2)/4$ and its projection along the $z$ direction in flavor space $T_z = m_A = n_A/2$, but there resides uncertainty on the other directions $\Delta T_x \sim \Delta T_y \sim \sqrt{n_A}$, which can be mimicked by a perturbation of $\Delta T_x/T_z \sim 1/\sqrt{N}$ in the mean-field picture.

\begin{figure}[t]
	\centering
		\includegraphics[width=0.48\textwidth]{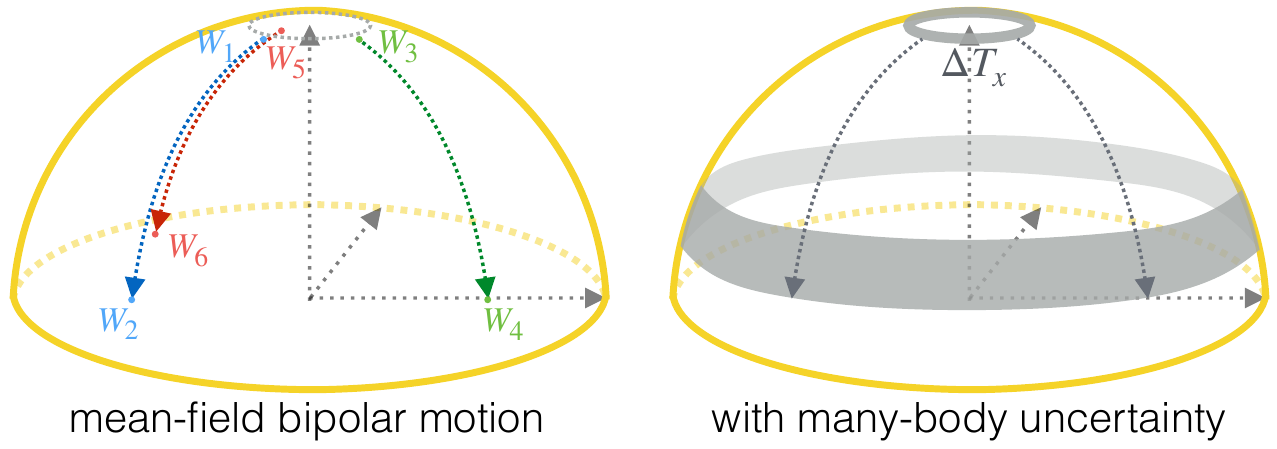}
    \caption{\label{fig:decoherence} Schematic diagrams for bipolar motions in the upper hemisphere of flavor isospin at mean-field approximation and with many-body uncertainty. On the left panel, the mean-field flavor isospin starting from point $W_1$, $W_3$, or $W_5$, will end up being at point $W_2$, $W_4$, or $W_6$ respectively. On the right panel, the many-body uncertainty near the top (upper grey ring) leads to the decoherence in both azimuthal and zenith directions (lower grey ring).}
\end{figure}

This uncertainty is not captured by the mean-field solution because the mean-field flavor isospin has a well-defined orientation characterized by a three-component vector analogous to a pendulum~\cite{duan2007picture,raffelt2007self}.
Once the pendulum gets perturbed from pointing to the top of the flavor isospin sphere, it undergoes bipolar motion periodically by swing down and returning back to the original level in each cycle.
A small shift on the azimuthal angle of the initial state for the pendulum from $W_1$ to $W_3$ in Fig.~\ref{fig:decoherence} can lead to completely different trajectories near the equatorial region.
Moreover, given the exponential increasing essence of bipolar motion in the linear regime, a small differentiation on the zenith angle between $W_1$ and $W_5$ can be enhanced significantly from high to low latitude.

The quantum counterpart of flavor isospin can spread over more orientations and many-body states so that the system does not undergo the classical bipolar motion as depicted by the archetypal pendulum.
Although the uncertainty of $\Delta T_x/T_z$ in the initial state is suppressed with increasing particle number $N$, it can result in a huge uncertainty with respect to the azimuthal angle in flavor space especially when the state is closer to the top of the flavor isospin sphere.
It also results in an uncertainty for the zenith angle of the relative orientation between the flavor isospin in beam $A$ and that in beam $B$.
Therefore, the evolution of many-body state does not follow any specific trajectory in mean-field approximation but in a more spreading and entangled way.
This spreading purely originated from the many-body aspect leads to the deviation of flavor evolution and eventually the decoherence.

\section{Summary and outlook}
We have developed a numerical scheme to implement the slow flavor instability in a two-beam model with zero vacuum mixing.
This method can be generalized in the applications with several more angular beams for fast flavor conversion \cite{roggero2022entanglement} or including a nonzero vacuum mixing in the case of matter-neutrino resonance.
It can be inferred that the total number of many-body bases goes dominantly in a power law, $N_\text{bases} \sim (N/N_\mathrm{beam})^{N_\mathrm{beam}-1}$, of the number of beams $N_{\text{beam}}$.
A non-zero vacuum mixing brings in an extra degree of freedom and effectively increases $N_\text{beam}$ by one.
For $N_\text{bases}=10^7$ with a zero vacuum mixing angle, the total number of neutrinos in models with three beams, four beams, and five beams are $\sim 10^4$, 900, and 300, respectively, much higher than the system of $\approx 23$ neutrinos that has the same size of state space.

This many-body decoherence is sensitive to assumptions on the initial states.
A nonzero vacuum mixing can attenuate the distinction between the mean-field and many-body solutions when the uncertainty of flavor isospin $1/\sqrt{N}$ is less than the angle of initial state to the mass eigenstate \cite{martin2021classical}.
In this case, the azimuthal uncertainty goes to zero in the limit of infinite flavor isospins, and this neutrino quantum system reduces back to classical.
On the other hand, this attenuation also depends on the specific astrophysical conditions such as the effective mixing angle in the presence of matter density, the number of neutrinos overlapping in wave packets, and collisions that relax them to approach back to flavor eigenstate.
In addition, the initial state we use in this paper is a mean-field state, but neutrinos that propagate in almost the same velocity can have a relatively longer time to overlap on the wave packets and build up entanglement even before reaching into a region where the flavor instability can be triggered.
The preexisting entanglement potentially promotes many-body decoherence and changes the outcome from mean-field prescription.
Although the consequence for a more realistic neutrino energy and angular distribution remains uncertain and requires more following studies, any nonvanishing pair correlations and entanglements due to this decoherence may bring in non-negligible corrections to the neutrino flavor content in astrophysical environments.

The many-body effect for neutrino oscillations can be potentially linked to other physics systems with dynamic phase transition.
Similar features of bipolar evolution and a rapid decoherence associated with the quantum pendulum in our results have also been observed in a spin-1 Bose-Einstein condensate~\cite{gerving2012non}.
These resemblances indicate a fundamental connection in physics and open the possibility of studying collective neutrino flavor phenomena in the laboratory.

\begin{acknowledgments}
This work was supported by the European Research Council (ERC) under the European Union's Horizon 2020 research and innovation program (ERC Advanced Grant KILONOVA No.~885281).
I thank Ermal Rrapaj, Alessandro Roggero, Meng-Ru Wu, Huaiyu Duan, and Yong-Zhong Qian for the many valuable discussions.
I thank Gabriel Mart\'inez-Pinedo for his suggestions and improvements on the final version of the manuscript.
I also thank the anonymous referee for constructive criticisms and helpful suggestions.
\end{acknowledgments}

\vspace{0.4in}

\textit{Note added}.---A two-beam model including non-zero flavor mixing is recently reported in Ref.~\cite{martin2021classical}. While they find that in the limit of infinite flavor isospins the mean-field and many-body solutions from an initial mean-field state (product state) coincide for simple observables, they also observe deviations when the mixing angle and system size meet certain requirements.

\end{document}